\documentclass{optica-article}

\journal{opticajournal} 

\articletype{Research Article}

\usepackage{lineno}
\usepackage{siunitx}%
\usepackage{graphicx}
\usepackage{dcolumn}
\usepackage{bm}

\usepackage[utf8]{inputenc}
\usepackage[T1]{fontenc}
\usepackage{mathptmx}
\usepackage{etoolbox}

\begin{document}

\title{Source-device-independent monolithically integrated QRNG in a black box with generation rate in excess of 30 Gbit/s}

\author{Peter Seigo Kincaid\authormark{1,*}, Lorenzo De Marinis\authormark{1}, Francesco Testa\authormark{1}, Nicola Andriolli\authormark{2}, and Giampiero Contestabile\authormark{1}}

\address{\authormark{1}TeCIP Institute, Scuola Superiore Sant'Anna, Via Moruzzi 1, 56124, Pisa, Italy\\
\authormark{2}Department of Information Engineering, University of Pisa, Via Caruso 16, 56122, Pisa, Italy
}

\email{\authormark{*}peterseigo.kincaid@santannapisa.it} 


\begin{abstract*} 
Quantum mechanics provides a secure means of generating random numbers, with applications in fields spanning scientific simulation to cryptography. The first source-device-independent monolithically integrated quantum random number generator is reported. With a generation rate of 35 Gbit/s, the device is based on an InP photonic integrated circuit with a quantum vacuum state entropy source, sampled by heterodyne coherent detection using an optical local oscillator. The entire device is conveniently housed in a black box and includes all the necessary driving and signal conditioning electronics, with electrical interfaces only; the exhibited security, compactness, and fast generation rate make the generator suitable for applications in QKD.
\end{abstract*}

\section{\label{sec:Introduction}Introduction}
The generation of random numbers is integral to many applications, including key generation in cryptography, scientific simulations and fundamental science~\cite{RevModPhys.89.015004}. Pseudorandom numbers can be generated algorithmically from a starting seed; however, given the same seed the sequence is reproducible, and the random numbers exhibit a strong long-range correlation~\cite{ma2016quantum}, which presents a vulnerability for cryptographic applications. On the other hand, true random number generators (TRNG) are based on unpredictable processes, known as entropy sources, such as thermal noise in electronics~\cite{MurryGeneral1970}, with a more recent implementation based on random telegraph noise in memristors~\cite{SongOptim_2023}. Quantum processes constitute a good entropy source for TRNG and compared to those based on classical effects, they are theoretically non-deterministic.

Quantum random number generators (QRNG) can be classified into different categories based on the guaranteed level of security of the device. The highest level of security is offered by device-independent (DI) QRNG, which are self-testing and where no assumptions on the generator are required; in this case, the randomness is verified by observing violations of the Bell inequalities~\cite{ma2016quantum}. These systems tend to have low generation rates, limiting their practical use~\cite{HighSpeedDILiu2018}. On the other hand, fully trusted systems offer the highest generation rates~\cite{100-GbitBruynsteen2023}, with the assumption that both the quantum source and measurement device are well characterized; however, when these assumptions are broken, e. g., by adversary manipulation of the device, the randomness could be compromised. A practical trade-off between generation rate and security is to bound assumptions to only a part of the QRNG, for example, on the quantum entropy source~\cite{Cao_2015} or the measurement device~\cite{qrng2018}, and assume that it is well-characterized, while making no assumptions about the rest of the system. This class of devices is defined as semi-device-independent (semi-DI) QRNG, and can address applications where a reasonable level of security and fast generation rate is critical (note that for a system with a 10 GHz clock a 40 Gbit/s random number generation rate would be required~\cite{Nie_68gps_2015,Takesue:06}).

Many QRNG schemes are based on bulky set-ups, with generation based on quantum fluctuations in photonic systems. Integrated photonics provides a way of miniaturizing these systems, with III-V materials such as Indium Phosphide (InP) offering the highest level of integration and compactness through the availability of coherent light sources in a monolithic fabrication platform. Few examples of fully monolithically integrated QRNG have been reported, with some bare chip demonstrations based on the interference of CW and gain-switched lasers~\cite{Abellan:16,chrysostomidis2023long}.
Very recently, fully packaged monolithically integrated QRNGs with real-time number extraction based on gain-switched pulse interference~\cite{roger2019real,marangon2024fast} and self-interference of a laser in an interferometric structure~\cite{Smith:25} have been reported. The increasing interest for compact and fast QRNGs is also demonstrated by another recent report of hybrid integration in package of an InP laser with a silicon-based balanced homodyne detection~\cite{hua2025fully}. These 
integrated systems are all fully trusted (device dependent) QRNGs exhibiting generation rates under 10 Gbit/s. Semi-DI QRNGs have a more complex design and are harder to integrate, for example, the system proposed in~\cite{qrng2018} was only partially integrated using silicon photonics, and relied on an external laser, VOA and power meter~\cite{Bertapelle:25,Bertapelle:25_conf} for the number generation, attaining generation rates of 17 and 20.2 Gbit/s respectively for the bulk and partially integrated versions. Simulations regarding a monolithic implementation of this QRNG were reported in a recent conference contribution~\cite{KincaidSourceDevice2023}. In this work, we report the full integration and packaging of this system on a monolithic compact InP chip demonstrating for the first time a compact and practical ``black box'' realization of a source-DI QRNG with no required optical I/O. To the author's knowledge, the reported generation results make it the current fastest QRNG in the semi-DI class.

\section{InP Monolithic PIC Design and Packaging}

\begin{figure}
\centering
\includegraphics[width=\textwidth]{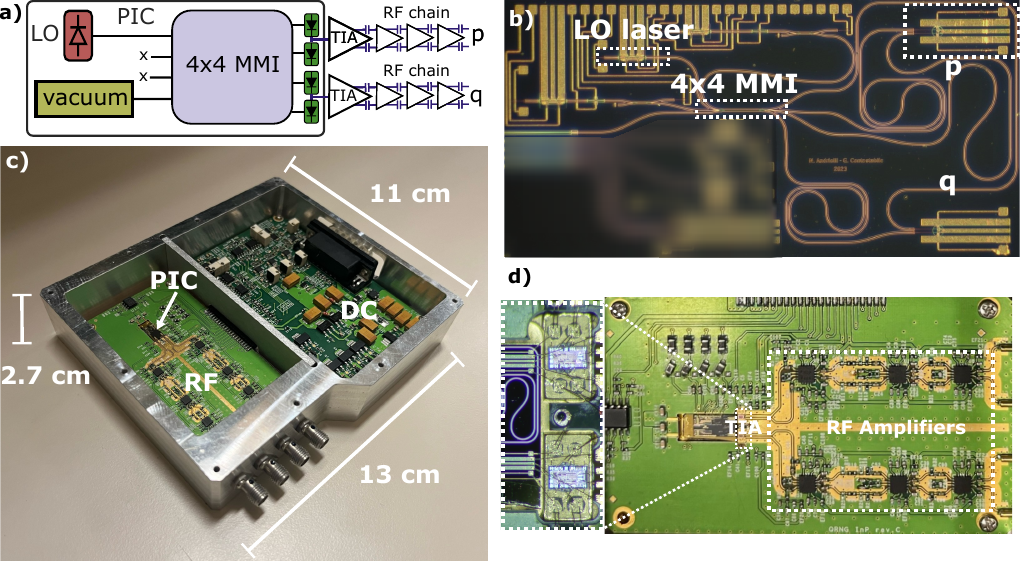}
\caption{a) Schematic: InP monolithic heterodyne-based QRNG with electrical amplification chain. b) Photo of the fabricated InP chip. c) Single module QRNG package. d) RF amplification chain and close-up of the wirebonded BPDs and TIAs.}
\label{fig:photos}
\end{figure}

The PIC was realized as part of a HHI dedicated wafer run, it has an active area footprint of around 24 mm$^2$ and is displayed in Fig.~\ref{fig:photos}b. The photonic circuit, whose schematic is presented in Fig.~\ref{fig:photos}a, consists of a DFB laser centered at 1550~nm acting as local oscillator (LO) for the input vacuum state, both inputs are connected to a 90$^{\circ}$ optical hybrid, implemented by a 4x4 multi-mode interference coupler with less than 0.06~dB simulated power imbalance between I and Q outputs. The 4 output waveguides are then sent via matched optical pathways into two balanced photodiodes which read the $p$ and $q$ quadratures of the quantum ground state.

The chip was mounted on an RF PCB, used for signal conditioning, which includes an electronic amplification chain similar to that in~\cite{Bertapelle:25}. The balanced photodiodes, having an estimated responsivity of 0.8~A/W at 1550 nm, are interfaced with low-noise transimpedance amplifiers (TIAs, Analog Devices ADN 2880) via ball-wedge wirebonds, displayed in Fig.~\ref{fig:photos}d. Here, the bandwidth was limited by the TIA (3 dB point at 2.5~GHz) and not by the parasitic capacitance of the wirebonds. These are connected to a multi-stage amplification chain composed of three differential amplifiers (Analog Devices ADA 4960) (Figs.~\ref{fig:photos}a, d). The resulting transimpedance of the whole amplification chain is 281.6 k\si{\ohm}. The RF board is interfaced with another DC control board (both custom made) that is used to operate the laser and the balanced detectors and to provide and readout circuit currents. The whole system is enclosed inside an aluminium box to protect the PIC and shield against EMI, with DC electrical input/output provided through a multi-pin cable, and RF output from the two differential electrical signals via SMA connectors (photo in Fig.\ref{fig:photos}c). The resulting device is a black box photonic-based QRNG with only electrical I/O. One of the two realized prototypes included a Peltier cooler, though since the aluminium case and PCB metallic base are inherently dissipative, the performance of the generator was not observed to change under Peltier operation.
The onboard DFB laser power is 13~mW and the received power at each single photodiode is 140~\si{\micro\watt}.
Thus, a propagation loss of 14~dB was estimated compared to the 4.5~dB expected from the PDK, this $\approx$~10~dB in additional losses negatively affects overall QRNG performance, as discussed in the following.


\section{Results}
\begin{figure}
\centering
\includegraphics[width=\textwidth]{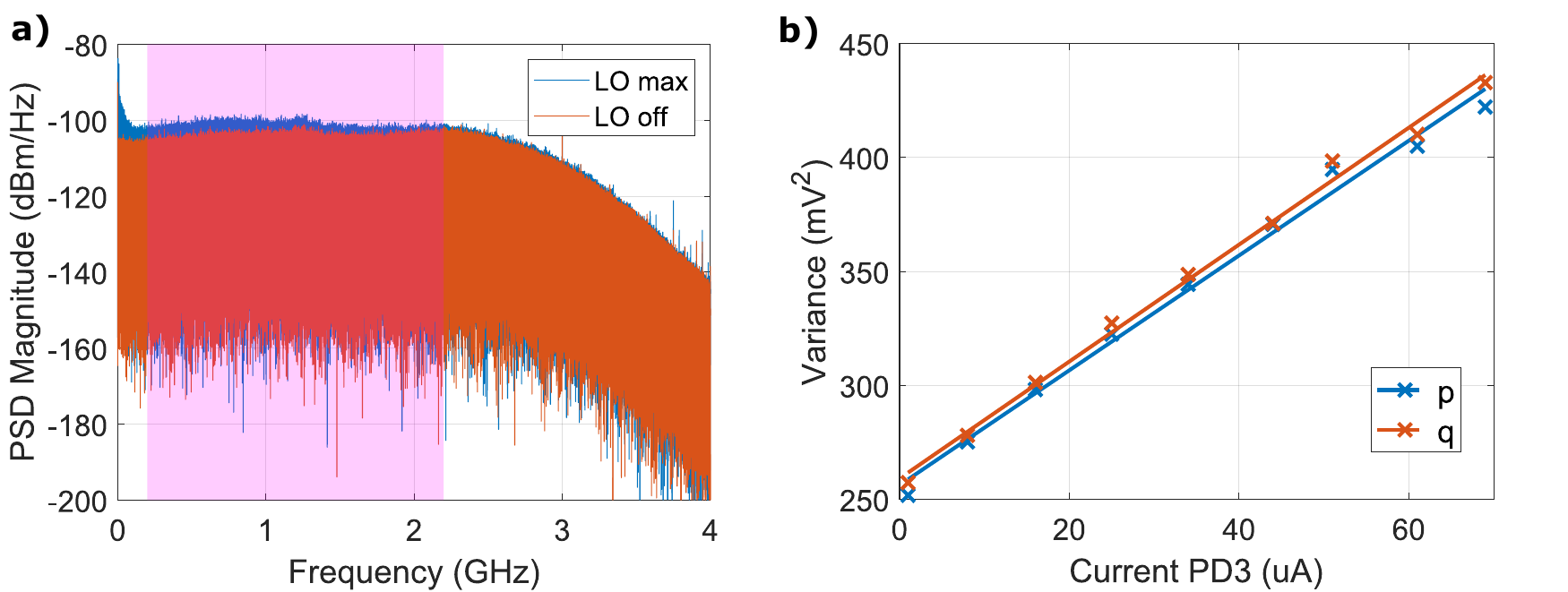}
\caption{a) Spectra of the p quadrature RF output, showing the difference when the LO is switched on at maximum power. b) Calibration procedure, variance of quadrature signals as a function of photodiode current.}
\label{fig:fig1}
\end{figure}

\begin{figure}
\centering
\includegraphics[width=\textwidth]{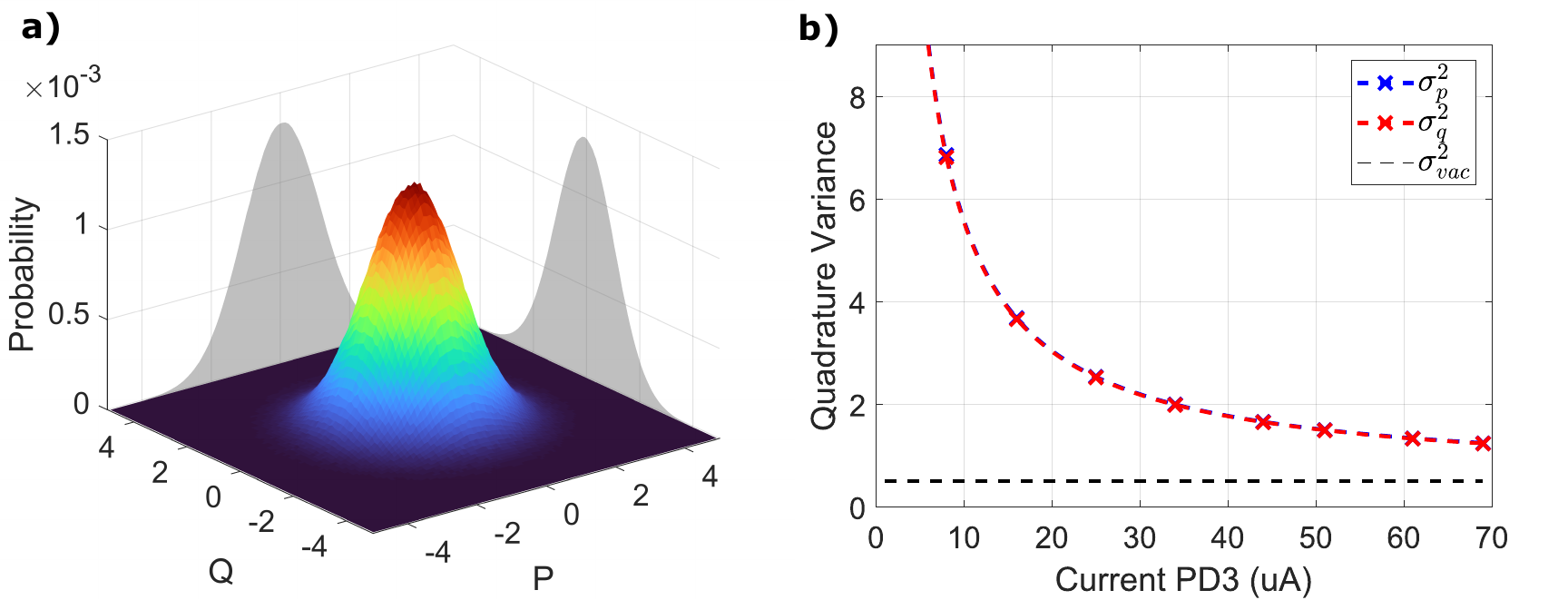}
\caption{a) Probability distribution of the received signals at both quadratures. The projections in the $P$ and $Q$ plane illustrate the individual shape of the $P$ and $Q$ quadratures, with height normalized to the 3D plot. b) Quadrature variance in vacuum units as a function of received photodiode current, expected vacuum quadrature variance shown for reference. }
\label{fig:fig3}
\end{figure}

One of each pair of RF outputs of the module was terminated with a 50~\si{\ohm} resistance in order to interface the device with a single-ended ADC, here a 12 bit oscilloscope with 20 GS/s was used for readout. The data were then transferred to a computer for off-line digital processing.

\begin{figure}
\centering
\captionsetup{width=0.9\textwidth}
\includegraphics[width=\textwidth]{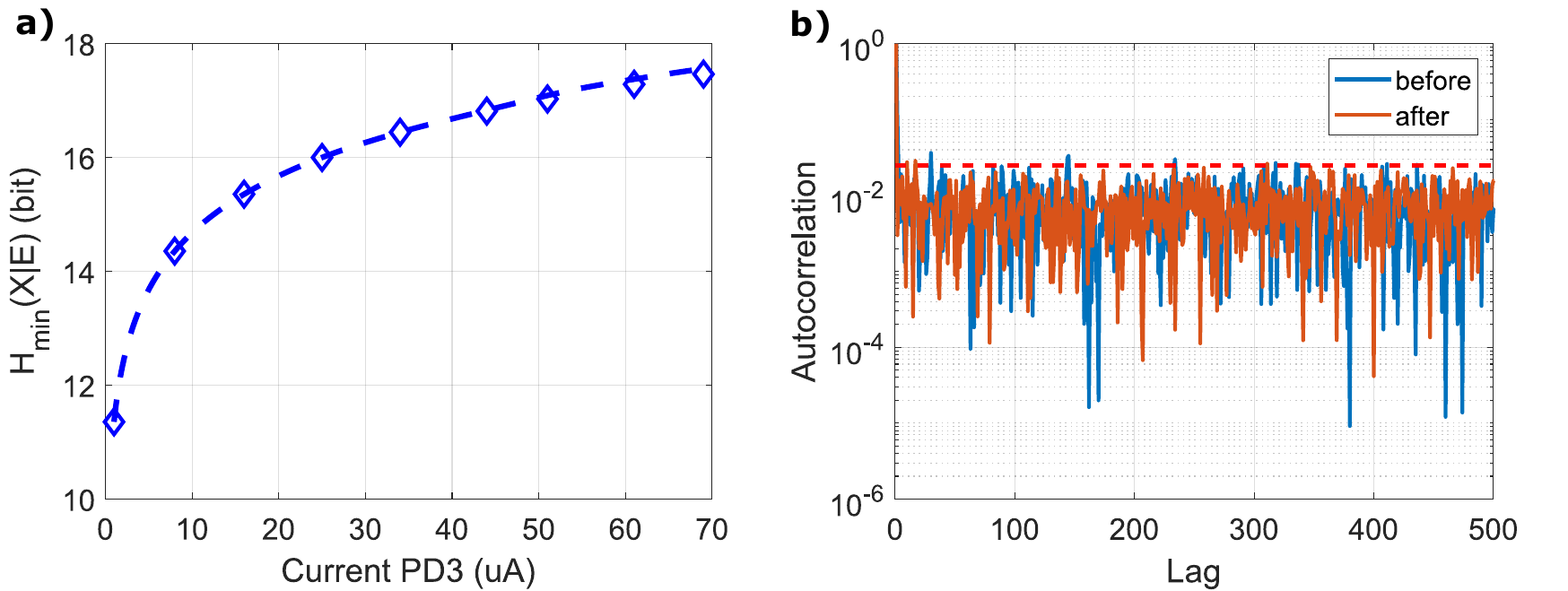}
\caption{a) Conditional quantum minimum entropy as a function of received photodiode current. b) Sample autocorrelation before and after hashing for a sample length of 10788.}
\label{fig:fig2}
\end{figure}

The resulting power spectral density at the RF output is displayed in Fig.~\ref{fig:fig1}a both for when the laser is off, and when the laser is at maximum power. Additional low frequency spectral noise is observed when the laser is switched on, this is eliminated with a band pass filter, applied digitally with 0.2-2.2~GHz passband (displayed in magenta in Fig.~\ref{fig:fig1}a), with the upper frequency cutoff lower than the 3-dB bandwidth of the TIA. The data are then downsampled at 2 GS/s to match the first zero of the autocorrelation function~\cite{qrng2018}. A calibration measurement was performed to estimate $H_{min}\left(X|E\right)$ by varying the optical power of the laser. The variance of the $p$ and $q$ quadratures is plotted as a function of the resultant DC photocurrent read at $PD3$ in Fig.~\ref{fig:fig1}b, the bottom photodiode of the $q$ quadrature from Fig.~\ref{fig:photos}a, as the laser power is increased to around 13~mW. The aim of this calibration procedure is to convert the ADC resolution from physical units to vacuum units, retaining resilience to differences in readout from the two quadratures~\cite{qrng2018}. The correction factor is given by $k_{p,q}=\sqrt{2 m_{p,q} I_{PD3}}$, where $m_{p,q}$ is the angular coefficient of the fitted curves in Fig.~\ref{fig:fig1}b, this is normally expressed in terms of power, but the correction factor is power-independent and here it is expressed in terms of measured photocurrent. Using this conversion, the probability distribution of the quadratures $p$ and $q$ are plotted in Fig.~\ref{fig:fig3}a in vacuum units from experimental data acquired at maximum laser power. The projections along the $p$ and $q$ axis show 2D slices of the measured distributions. The measured variance at both quadratures in vacuum units is plotted in Fig.~\ref{fig:fig3}b, where the reference level corresponding to $\sigma_{vac}^2$=0.5 is also displayed. The minimum variance at maximum laser power is 1.23 for both quadratures, higher than $\sigma_{vac}^2$ due to the presence of classical noise and reflected by a wider cone than expected from the Husimi function in Fig.~\ref{fig:fig3}a. This impurity of the quantum state is taken into account by the randomness extraction as later discussed. The resolution in vacuum units is given by $\delta_{p,q}=R_{adc}$/$2^{N}k_{p,q}$, where $R_{adc}$ is the full scale range of the ADC and $N$ is the resolution in bits. The appendix reports the way the quantum minimum entropy, the number of extractable random bits based on the quantum side information available to an attacker, $H_{min}\left(X|E\right)$, is related to $\delta_{p,q}$. Fig.~\ref{fig:fig2}a displays $H_{min}\left(X|E\right)$ as a function of the DC current received at the photodiode; at the maximum laser power, $H_{min}\left(X|E\right)$~=~17.5 bits; this is larger than the minimum entropy reported in~\cite{Bertapelle:25} due principally to the ADC we used which had a higher bit resolution, and despite a lower ratio of $H_{min}\left(X|E\right)$/$2N$ in our case due to the limited power arriving at the BPDs and higher propagation losses in the InP PIC. In our case, with a wider passband the quantum state is less pure, due to a variable clearance, this results in a larger deviation from $\sigma_{vac}^2$; despite this increased impurity, since $\delta_{p,q}\propto1$/$\sqrt{m_{p,q}}$ and $m_{p,q}$ increases with a wider passband, the resolution in vacuum units decreases, leading to an increased $H_{min}$.

After the calibration procedure, the LO laser power remained at maximum, for which the current at PD3 was 70~\si{\micro\ampere} and data were acquired at the ADC and saved offline. The filtered and downsampled data were processed using the procedure detailed in~\cite{ma2013postprocessing} by dividing the data into blocks and multiplying each by a Toeplitz matrix of dimension, 10788$\times$15000, for which the condition number of rows/number of columns < $H_{min}\left(X|E\right)$/$/2N$ must be satisfied and for our case yields a security parameter, $\epsilon<2^{-63}$. From 15~GB of raw unfiltered data, we had a final processed data size of around 1~GB. Fig.~\ref{fig:fig2}b presents a comparison in the absolute value of the autocorrelation, normalized by the sample variance, before and after randomness extraction for a sample length of 10788. The 99$\%$-confidence interval for a Gaussian white noise process is plotted for reference as a red dashed line~\cite{Smith:25}, showing the boundary under which 99$\%$ of all autocorrelation values should lie. Before the hashing for the second lag the value is around 4 times larger than the 99$\%$ boundary, likely due to noise in our sampling equipment, as observed in~\cite{qrng2018}. The data after processing do not exhibit this effect and are uncorrelated. As a double check that our post-processing procedure was correct, and that there weren't any errors in data acquisition, the NIST SP 800-22 battery of statistical tests was implemented with results reported in table~\ref{tab:NIST}. The 1~GB of postprocessed data were tested by dividing into 8000 bit streams of 10$^6$ length, and passed all the statistical tests. With our rate of downsampling and $H_{min}$, the resulting generation rate of the QRNG is 35~Gbit/s. Indeed, using a downsampling frequency of double the band pass frequency as in~\cite{Bertapelle:25}, we would obtain a 70~Gbit/s generation rate, though we observed a strong negative autocorrelation between adjacent data points as predicted by the Wiener-Khinchin theorem.
Additionally, we estimate that, with optical losses as expected from the foundry PDK, we would have approximately 10~dB of additional optical power at the BPDs. Since the variance of the quantum noise scales linearly with the received optical power, the resolution at each quadrature in vacuum units, $\delta_{p,q}$, would be reduced by a factor of $\sqrt{10}$ and the resulting $H_{min}\left(X|E\right)$ would be increased to 20.8 leading to a generation rate of 41.6~Gbit/s.

\begin{table}[h!]
  \centering

\begin{tabular}{c c c}

\textbf{Statistical Test} & \textbf{P-value} & \textbf{Result} \\
\hline
Frequency                      & 0.1962 & PASSED \\
BlockFrequency                 & 0.4805 & PASSED \\
CumulativeSums                 & 0.3181 & PASSED \\
Runs                           & 0.5119 & PASSED \\
LongestRun                     & 0.4466 & PASSED \\
Rank                           & 0.1194 & PASSED \\
DFT                            & 0.1540 & PASSED \\
NonOverlappingTemplate         & 0.0085 & PASSED \\
OverlappingTemplate            & 0.1412 & PASSED \\
Universal                      & 0.2306 & PASSED \\
ApproximateEntropy             & 0.5631 & PASSED \\
RandomExcursions               & 0.1722 & PASSED \\
RandomExcursionsVariant        & 0.0232 & PASSED \\
Serial                         & 0.7134 & PASSED \\
LinearComplexity               & 0.6944 & PASSED \\
\hline
\end{tabular}

\caption{NIST test results for a bit stream length of 10$^6$ with 8000 bit streams (approx 1GB of random bits). P-value represents the uniformity of individual p-values, they are considered non-uniform if P-value<0.0001. When more than one test is performed for a given test category, the lowest P-value is reported in the table.}
\label{tab:NIST}
\end{table}

\section{Discussion and Conclusions}
We reported for the first time a source-device-independent QRNG in a black-box with 35 Gbit/s generation rate capability. The device leverages a monolithically integrated InP circuit and includes all the driving and analog signal conditioning electronics.
As a general note on generation rate, it does not depend solely on PIC performance, but also on the following signal conditioning and processing stages, i.e., on the specific ADC, TIAs, amplification stages and frequency filtering used. In our case, we performed analog data acquisition with a 12-bit ADC with 20~GSample/s capacity, the use of ADCs with lower resolution would lead to lower generation rates. This would likely be the case for a real-time number generator implementation, if made using relatively low-cost commercial ADCs and FPGAs, though generation rates in the multi Gbit/s range should still be easily attainable. For this reason, the pure comparison of achieved generation capacity among different schemes or implementations has limited significance in real applications. Nevertheless, in this manuscript we demonstrate that a source-DI-QRNG can be realized based on a self-consistent monolithically integrated PIC (i.e., a PIC including the photonic entropy source and the detection block) such that it can be mounted in a compact black-box with only electrical I/O.
This is a valuable result in the direction of implementing practical and compact systems which need secure quantum numbers at high rate, including those required by QKD.

\section*{Appendix}
The source independence of the heterodyne measurement lies in the ability to mathematically describe the minimum quantum entropy in the case when the quantum source is completely controlled by the attacker, Eve and the presence of quantum side information available to them, $\epsilon$. The idea is that the attacker exploits the information $\epsilon$, knowledge of the set of positive operator value measurements (POVM) performed, in order to maximise the probability of guessing the outcome $X$, which can be expressed as~\cite{qrng2018,Bertapelle:25}:

\begin{equation}
    P_{guess}\left(X|\epsilon\right)\leq\text{max } \text{Tr}\left[\hat\Pi_A\hat\tau_A\right],
    \label{eq:prob}
\end{equation}
where $\hat\Pi_A$ represents the discretized POVM corresponding to the heterodyne measurement, and $\tau_A$ the superposition of states used by Eve. From the Leftover Hash Lemma~\cite{RennerSecurity2008}, the number of extractable bits in the presence of quantum side information is bounded by:

\begin{equation}
    H_{min}\left(X|\epsilon\right)=-\text{log}_2P_{guess}\left(X|\epsilon\right),
\end{equation}
hence from Eq.\ref{eq:prob}\cite{qrng2018}:
\begin{equation}
    H_{min}\left(X|\epsilon\right)\geq\text{log}_2 \frac{\pi}{\delta_p\delta_q},
    \label{eq:Hmin}
\end{equation}
where $\delta_{p,q}$ are the resolutions of the $p$ and $q$ quadratures in vacuum units. Using this formula, the randomness can be extracted using a universal hashing method based on Toeplitz matrices~\cite{ma2013postprocessing}.

\begin{backmatter}
\bmsection{Funding}
The authors acknowledge the financial support provided by the National Quantum Science and Technology Institute (NQSTI) through the PNRR MUR project PE0000023-NQSTI and by the Italian MUR in the framework of the FoReLab project (Departments of Excellence). 

\bmsection{Acknowledgments}
The authors thank Davide Rotta and Marco Chiesa from CamGraPhIC for their valuable expertise in packaging, and Alberto Montanaro (CNIT) and Alberto Santamato (Nu Quantum Ltd.) for their insightful and fruitful discussions.

\bmsection{Disclosures}
The authors declare no conflicts of interest.

\bmsection{Data Availability Statement}
Data underlying the results presented in this paper are not publicly available at this time but may be obtained from the authors upon reasonable request.

\bigskip

\end{backmatter}

\bibliography{bibfile}






\end{document}